\documentclass[showpacs,twocolumn,aps,eqsecnum]{revtex4}
\usepackage[dvips,final]{graphicx}
\usepackage{epsfig}
\usepackage{amssymb}

\begin{document}
\title{Theory of biphoton generation in a single-resonant optical parametric oscillator\\
far below threshold}
\author{Ulrike Herzog, Matthias Scholz, and Oliver Benson}
\affiliation{Nano-Optics, Institut f\"ur Physik, Humboldt-Universit\"at zu Berlin, D-10117 Berlin, Germany}
\date{\today}
\begin{abstract}
We present a quantum-theoretical treatment of biphoton generation in single-resonant type-II parametric down-conversion. The nonlinear medium is
continuously pumped and is placed inside a cavity which is resonant for the signal field, but nonresonant for the idler deflected by an intra-cavity
polarizing beam splitter. The intensity of the classical pump is assumed to be sufficiently low in order to yield a biphoton production rate that is
small compared to the cavity loss rate. Explicit expressions are derived for the rate of biphoton generation and for the biphoton wave function. The
output spectra of the signal and idler field are determined, as well as the second-order signal-idler cross-correlation function which is shown to be
asymmetric with respect to the time delay. Due to frequency entanglement in the signal-idler photon pair, the idler spectrum is found to reveal the
longitudinal mode structure of the cavity, even though the idler field is not resonant.
\end{abstract}
\pacs{03.67.-a,42.50.-p,42.50.Ar,42.65.Lm}
\maketitle

\section{Introduction}
In parametric down-conversion, a pump photon of frequency $\omega_p$ incident on a medium with a second-order nonlinear susceptibility $\chi$ is
split into two photons with lower frequency~\cite{mandel}. Spontaneous parametric down conversion produces photon pairs that can be entangled in many
degrees of freedom~\cite{kwiat}. The resulting two-photon state, consisting of a signal photon at frequency $\omega_s$ and an idler photon at
frequeny $\omega_i= \omega_p-\omega_s$, is often called a biphoton. Upon post-selection on the idler photons, the process provides a source of
heralded single photons which represent the principal resource in many quantum information processing protocols like quantum
cryptography~\cite{1,2,3} or linear optics quantum computation~\cite{4}. Quantum networks have been proposed~\cite{5,6} that rely on stationary atoms
or ions as information processing nodes and on single photons to transmit information via optical fibers. First building blocks of this scheme have
already been realized~\cite{Felinto,Rempe}. For an efficient atom-photon coupling, the photon bandwidth has to match the linewidth of the atomic
transition which is by orders of magnitude smaller than the bandwidth of the photons emitted in spontaneous parametric down-conversion.

In order to reduce the photon bandwidth, cavity-enhanced parametric down-conversion can be applied. A bright source of heralded narrow-band single
photons was experimentally realized~\cite{polzik} using a double-resonant optical parametric oscillator (OPO). The nonlinear crystal was placed
inside a cavity resonant for both the signal and the idler field, and in the output a single narrow-band longitudinal signal mode was selected with
an external frequency filter. Clearly, to conditionally achieve single-photon generation, the OPO has to be operated in the regime far below
threshold where the production rate of down-converted photons is small compared to the loss rate of the cavity. A number of preceding experiments for
biphoton generation in a double-resonant OPO far below threshold have been performed~\cite{lu,lu1,lu2,wang,kukl1,kukl2}, and the first theoretical
description was given in Ref.~\cite{lu1}. Recently, the theory of the double-resonant OPO has been extended by analyzing the conditionally prepared
single-photon state~\cite{nielsen1}, and by providing a multi-mode treatment which is valid for both pulsed and stationary pump
fields~\cite{nielsen2}.

To ensure reliable operation of a quantum network, continuous photon emission over a long period of time is essential. For this purpose, an active
stabilization of the OPO is necessary which proves to be a complicated task in the double-resonant case while it is easier to achieve for a
single-resonant cavity. Continuous biphoton generation in a single-resonant OPO far below threshold has recently been demonstrated in our
group~\cite{scholz} using a setup where only the signal mode experiences resonance enhancement in the cavity while the orthogonally polarized idler
mode is nonresonant due to deflection by an intra-cavity polarizing beam splitter. Additional passive filtering with the help of an external cavity
can select a single longitudinal mode and will thus enable the generation of narrow-band single photons.

To our knowledge, a theoretical treatment of the single-resonant OPO far below threshold has not been performed so far. The present paper aims to
fill this gap. Based on the concepts of the pioneering theoretical studies of spontaneous parametric down-conversion~\cite{hong,ghosh}, we provide
the theoretical background for our experimental results~\cite{scholz}. Different from the approaches used for the theoretical description of the
double-resonant OPO~\cite{lu1,nielsen1,nielsen2}, free-field quantization of the idler field is inevitable for our scheme.

This paper is organized as follows: In Sec.~II, the basic equations for describing the nonlinear interaction between the quantized signal and idler
fields are provided. The biphoton production rate and the biphoton wave functions are derived in Sec.~III by applying the standard perturbative
treatment in the Schr\"odinger picture~\cite{mandel}. The results are used in Sec.~IV to derive the spectral properties of the emitted radiation and
to study the the second-order signal-idler cross-correlation function. Sec.~V concludes the paper by establishing the connection to real experimental
situations.

\section{Basic equations}
\subsection{The interaction Hamiltonian}
We consider type-II parametric down-conversion in a nonlinear crystal of length $l$ that is pumped by a monochromatic linearly polarized classical
field of frequency $\omega_P$. The crystal is assumed to be placed inside a cavity which is resonant for the signal field, but not for the idler
field, polarized orthogonal to the signal. Fig.~1 shows a schematic picture of the corresponding experimental setup~\cite{scholz}.
\begin{figure}[h!]
\center{\includegraphics[scale=0.6,draft=false]{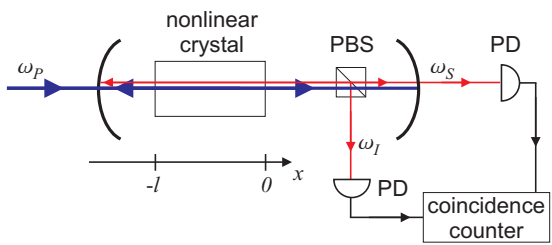}} \caption{(Color online) Scheme of the considered experiment. The cavity acts as a
one-sided resonator for the signal field, while the idler field is non-resonant due to deflection at an intra-cavity polarizing beam splitter. The
signal-idler cross-correlation function  is determined by delayed coincidence detection.}
\end{figure}
The central frequencies $\omega_S$ and $\omega_I$ of the signal and idler field depend on the properties of the birefringent nonlinear crystal and
are determined by energy and momentum conservation known as phase-matching~\cite{mandel}
\begin{eqnarray}
\label{omega}
\omega_P & = & \omega_S + \omega_I,\\
\label{k} \vec{k}_p(\omega_P) & = & \vec{k}_s(\omega_S) + \vec{k}_i(\omega_I).
\end{eqnarray}
Here, $\vec{k}_p$, $\vec{k}_s$, and $\vec{k}_i$ are the wave vectors of the pump, signal, and idler waves which are collinear in the considered
setup, with the signal leaving the cavity in positive $x$-direction. The standing-wave fields of the pump and the signal inside the cavity are
composed of two components, propagating in negative and positive $x$-direction, respectively. Because of the phase-matching conditions, only those
components contribute to the parametric interaction that have the same propagation direction as the idler wave. Neglecting the vector notation of the
fields, the positive-frequency part of the electric field in the relevant component of the classical pump inside the crystal can be written as
\begin{equation}
\label{pump} E_{P,cr}(x,t)= E_P\,{\rm e}^{i[k_p(\omega_P) \,x -
\omega_P t]}.
 \end{equation}
The weak signal and idler fields inside the crystal are described by the operators $E_{S,cr}^{(+)}=  E_{S,cr}^{(-)\dag}$ and $E_{I,cr}^{(+)} =
E_{I,cr}^{(-)\dag}$, respectively, denoting the positive-frequency part of their co-propagating field components. The interaction Hamiltonian can be
written in the simplified form
\begin{equation}
\label{H-int} H_{int}=\frac{\chi}{2l}\int_{-l}^0 dx\,
( E_{P,cr}E_{S,cr}^{(-)}E_{I,cr}^{(-)} +E_{P,cr}^{\ast}E_{S,cr}^{(+)}E_{I,cr}^{(+)}),
\end{equation}
where the second-order nonlinear susceptibility $\chi$  is frequency-dependent~\cite{mandel}. Before utilizing Eq.~(\ref{H-int}), we need to find
explicit expressions for the operators $E_{S,cr}^{(+)}$ and $E_{I,cr}^{(+)}$.

\subsection{The free-field operators}
For later use, we start by providing the operators for the signal and idler fields in free space. The positive-frequency electric field operator of a
wave with transverse cross-section~$A$ propagating freely in $x$-direction is given by $ E^{(+)}(x,t) =
\lim_{L\to\infty}\sum_{j=0}^{\infty}\sqrt{\frac{\hbar\omega_j}{2\epsilon_0 L A}} a_j {\rm e}^{i\omega_j(\frac{x}{c} -  t)}$ with $\omega_j=2\pi j
c/L$ where $L$ is the quantization length and $a_j$ denotes the photon annihilation operator of mode~$j$. Let us consider the signal field and
introduce the frequency difference $\Omega_j= \omega_j-\omega_S$. Using $\Delta\Omega=2\pi c/L$, the transition to the continuum limit is performed
via the replacement $\sum_j a_j \ldots \rightarrow (\Delta \Omega)^{-1/2}\int_{-\omega_S}^{\infty} d\Omega a(\omega_S+\Omega)\ldots$ where the
continuous field operators $a(\omega)$ have the dimension ${\rm s}^{1/2}$. Since the bandwidth of the signal is small compared to its central
frequency $\omega_S$, the integration interval can be extended to $-\infty$, and we arrive at the approximate operator representation
\begin{equation}
\label{sig-out} E_S^{(+)}(x,t)=  \sqrt{\frac{\hbar\omega_S}{2\epsilon_0 c A}}
 \int_{-\infty}^{\infty}\frac{d\Omega}{\sqrt{2\pi}}\;
a(\omega_S+\Omega){\rm e}^{i(\omega_S + \Omega)(\frac{x}{c}-t)},\\
\end{equation}
where~\cite{walls}
\begin{equation}
\label{com-a} [a(\omega_1),a^{\dag}(\omega_2)]=\delta(\omega_1-\omega_2).
\end{equation}
Similarly, the corresponding operator for the idler field in free space is given by
\begin{equation}
\label{id1} E_I^{(+)}(x,t)= \sqrt{\frac{\hbar\omega_I}{2\epsilon_0 c A}} \int_{-\infty}^{\infty}
\frac{d\Omega}{\sqrt{2\pi}}\; b(\omega_I+\Omega){\rm e}^{i(\omega_I + \Omega)(\frac{x}{c} -  t)},\\
\end{equation}
where
\begin{eqnarray}
\label{comm1} [b(\omega_1),b^{\dag}(\omega_2)] =\delta(\omega_1-\omega_2).
\end{eqnarray}
Since in type-II parametric down-conversion signal and idler photons
are polarized orthogonally, we have
\begin{eqnarray}
\label{comm2} [a(\omega_1),b^{\dag}(\omega_2)] = 0.
\end{eqnarray}

\subsection{The field operators inside the crystal}
We now turn to the fields inside the crystal. If the nonlinear interaction is small, we do not need to consider the complicated problem of field
quantization in a nonlinear medium, but we can represent the field operators inside the crystal by adapting the corresponding expressions for the
free-field operators in order to account for the presence of a lossless dispersive medium~\cite{mandel}. In accordance with Ref.~\cite{blow}, the
operator for the positive-frequency part of the idler field inside the crystal then takes the approximate form
\begin{eqnarray}
\label{id-cr}\lefteqn{ E_{I,cr}^{(+)}(x,t)=}\\
&&\sqrt{\frac{\hbar\omega_I}{2\epsilon_0 c A n_I}}
\int_{-\infty}^{\infty} \frac{d\Omega}{\sqrt{2\pi}}\;
b(\omega_I+\Omega){\rm e}^{i[k_{I}(\Omega) x - (\omega_I+\Omega)
t]},\nonumber
\end{eqnarray}
where we introduced the wave vector at frequency $\omega_I + \Omega$,
\begin{equation}
\label{kI} k_{I}(\Omega)=\frac{\omega_I + \Omega}{c}\,n_i(\omega_I +\Omega).
\end{equation}
Here, the replacements $\epsilon_0\rightarrow \epsilon_0 n_i^2$ and $c \rightarrow c/n_i$ have been performed where $n_i$ is the refractive index of
the idler wave and $n_I = n_i(\omega_I)$.

To describe the signal field, we have to take the presence of the resonator into account. Let us first assume a lossless resonator completely filled
with the nonlinear medium. The quantization length of the field is then  equal to the crystal length $l$. When $n_s$ denotes the refractive index of
the signal, the adapted resonator eigenfrequencies characterizing the longitudinal modes can be written as
\begin{equation}
\label{long} \omega_m =\frac{(m_0+m)\pi c}{n_s (\omega_m)\;l} \quad
{\rm with}\quad m_0=\frac{\omega_S\,n_S\,l}{\pi c},
\end{equation}
where  $n_S = n_s(\omega_S)$,  $m=0, \pm1, \pm2, \ldots$ and $m_0\gg|m|$. Since the frequency difference between adjacent modes is small compared to
the total spectral width of the signal, as will become obvious in Sec.~IV, we can assume without lack of generality that $\omega_S$ coincides with
the frequency of a longitudinal mode and $m_0$ is an integer, i.e. that the cavity is tuned to resonance. Using the Taylor expansion
$n_s(\omega_m)=n_S+(\omega_m-\omega_S)\frac{\partial n_s}{\partial \omega}\left|_{\omega=\omega_S}\right.$, we find from Eq.~(\ref{long}) after minor
algebra that
\begin{equation}
\label{long1} \omega_m \approx  \omega_S + m \frac
{\pi\,v_{g,S}}{l}\equiv \omega_S +m \,\Delta\omega_c\, ,
\end{equation}
where
\begin{equation}
\label{vg} v_{g,S}=\frac{c}{n_S+\omega_S\frac{\partial n_s}{\partial \omega}\left|_{\omega=\omega_S}\right.}
\end{equation}
is the group velocity of the signal at frequency~$\omega_S$. By adapting the empty-cavity field operator~\cite{walls} with the replacements
$\epsilon_0\rightarrow \epsilon_0 n_s^2$ and $c \rightarrow c/n_s$, the part of the standing-wave signal field operator inside the crystal that
corresponds to a component traveling in positive $x$-direction in the lossless resonator is found to be
\begin{equation}
\label{sig0-cr}
E_{S,cr}^{(+)}(x,t)=\sqrt{\frac{\hbar\omega_S\Delta\omega_c}
{\epsilon_0 n_S c A \pi}}\sum_{m=-\infty}^{\infty}
 a_m\frac{{\rm e}^{i\omega_m [\frac{x}{c}n_s (\omega_m)- t)]}}{2}\,.
\end{equation}
Here, we replaced the quantization length under the square-root sign by the expression  $l= {\pi\,v_{g,S}}/{\Delta\omega_c}\approx {\pi\,c}/({n_S
\Delta\omega_c})$, following Eqs.~(\ref{long1}) and (\ref{vg}). Moreover, the summation has been extended to $m=-\infty$, in analogy to the expanded
integration range in Eq.~(\ref{sig-out}). The photon annihilation and creation operators for mode~$m$ obey the usual commutation relation $ [a_m
,a_{m^{\prime}}^{\dag}] = \delta_{m,m^{\prime}}$.

When resonator losses are incorporated, the modes turn into quasi-modes and the annihilation operators $a_m$ in Eq.~(\ref{sig0-cr}) become
time-dependent. According to the input-output formalism~\cite{collett,walls} for a one-sided cavity with loss constant $\gamma$, the damping of mode
$m$ is described by
\begin{eqnarray}
\label{in-out} \dot{a}_m(t) = -\frac{\gamma}{2}a_m(t) + \sqrt{\gamma} a_{m}^{IN}(t).
\end{eqnarray}
The operators $a_{m}^{IN}(t)$ and $a_{m}^{OUT}(t)$ characterize the ingoing and outgoing photon flux at frequency $\omega_m$ and have the dimension
s$^{-1/2}$. They are related by the boundary condition
\begin{eqnarray}
\label{in-out1}
 a_{m}^{IN}(t)= \sqrt{\gamma} a_m(t) - a_{m}^{OUT}(t).
\end{eqnarray}
In order to determine $a_m(t)$, we use the representation
\begin{equation}
\label{four} a_{m}^{OUT}(t)=  \frac{1}{\sqrt{2\pi}}
\int_{-\infty}^{\infty}d\Omega\,a(\omega_m+\Omega){\rm e}^{-i \Omega t},
\end{equation}
where, in analogy to Eq.~(\ref{com-a}),
\begin{eqnarray}
\label{comm3} [a(\omega_m+\Omega),a^{\dag}(\omega_{m^{\prime}}+\Omega^{\prime})]
=\delta_{m,m^{\prime}}\delta(\Omega-\Omega^{\prime}).
 \end{eqnarray}
Eq.~(\ref{comm3}) implies that the quasi-modes do not overlap which is justified in the good-cavity limit
\begin{equation}
\label{goodcavity}
 \gamma \ll \Delta\omega_c.
\end{equation}
After inserting Eq.~(\ref{in-out1}) into Eq.~(\ref{in-out}), we obtain by Fourier transformation the solution
\begin{equation}
\label{four1}  a_{m}(t)= \frac{1} {\sqrt{2\pi}}
\int_{-\infty}^{\infty}d\Omega \;a(\omega_m+\Omega)
\frac{\sqrt{\gamma}}{\frac{\gamma}{2}+i\Omega}\, {\rm e}^{-i \Omega
t}
\end{equation}
which has to be applied to Eq.~(\ref{sig0-cr}). The operator for the relevant field component of the signal in the lossy cavity can then be written
as
\begin{eqnarray}
\label{sig-cr} \lefteqn{E_{S,cr}^{(+)}(x,t)=\sqrt{\frac{\hbar\omega_S}
{2 \epsilon_0 n_S c A }}\frac{\sqrt{\gamma\Delta
\omega_c}}{2\pi}}\\
&&\times \sum_{m=-\infty}^{\infty}
 \int_{-\infty}^{\infty}
 d\Omega\;
\frac{a(\omega_m+\Omega)}{\frac{\gamma}{2}+i\Omega} \;
{\rm e}^{i[k_{S,m}(\Omega) x - (\omega_m+\Omega)t]}.\qquad\nonumber
\end{eqnarray}
Here,
\begin{equation}
\label{kS} k_{S,m}(\Omega)=\frac{\omega_m + \Omega}{c}n_s(\omega_m + \Omega)
\end{equation}
is the wave vector corresponding to a traveling-wave component of frequency $\omega_m + \Omega$. The denominator in the integral in
Eq.~(\ref{sig-cr}) describes radiation suppression for frequencies $\omega$ with $|\omega-\omega_m|=|\Omega|\gg \gamma$ while resonance enhancement
occurs for $\omega \approx \omega_m$.

So far, we have assumed a resonator length~$L_r$ that coincides with the crystal length~$l$. If $L_r>l$, a rigorous quantization of the signal field
has to account for the exact position of the crystal inside the resonator, but is beyond the scope of the present paper. For the purposes of our
approximative treatment, however, it is sufficient to describe the signal field inside the crystal by Eq.~(\ref{sig-cr}) with
$\Delta\omega_c=\Delta\omega$ and $\omega_m = \omega_S + m\Delta\omega$ where $\Delta\omega$ is the effective free spectral range. The latter can be
represented as
\begin{equation}
\label{T} \Delta\omega=\frac{2\pi}{T}\quad {\rm with}\quad T
=\frac{2l}{v_{g,S}}+\frac{2(L_r-l)}{c},
\end{equation}
where $T$ is the effective cavity round-trip time of a signal photon.

\section{The rate of biphoton generation and the biphoton wave function}
With the expressions for the operators of the signal and idler field at hand, we are now in the position to specify the interaction Hamiltonian and
to derive a perturbative solution of the Schr\"odinger equation. Making use of  Eqs.~(\ref{omega}), (\ref{pump}), (\ref{id-cr}) and (\ref{sig-cr}),
as well as Eq.~(\ref{long1}) with $\Delta\omega_c \rightarrow \Delta\omega$ and taking the frequency-dependence of the nonlinear susceptibility into
account, we find from Eq.~(\ref{H-int})
\begin{eqnarray} \label{H-int1} H_{int}=i \hbar \alpha
\sum_{m=-\infty}^{\infty} \int_{-\infty}^{\infty}d\Omega\, \frac{ \sqrt{\gamma}}{\frac{\gamma}{2}-i\Omega}
\int_{-\infty}^{\infty}d\Omega^{\prime}\,F_m
(\Omega,\Omega^{\prime})\nonumber\\
 \times \,a^{\dag}
(\omega_m+\Omega) b^{\dag} (\omega_I+\Omega^{\prime}) {\rm e}^{ i(m
\Delta\omega + \Omega + \Omega^{\prime})t} + H. A.,\quad
\end{eqnarray}
where we introduced the function
\begin{eqnarray} \label{F-m}
\lefteqn{F_m(\Omega,\Omega^{\prime})=}\\
&&\frac{\chi(\omega_P;\omega_m+\Omega,\omega_I+\Omega^{\prime})}
{\chi(\omega_P;\omega_S,\omega_I)\,l} \int_{-l}^0 dx\; {\rm e}\,^{i [k_P -
k_{S,m}(\Omega)-k_I(\Omega^{\prime}]}\nonumber
\end{eqnarray}
and defined the constant
\begin{eqnarray} \label{alpha}
\alpha= \frac{ - i E_P  }{8\pi\epsilon_0 c A}
\sqrt{\frac{\omega_S\omega_I}{n_S n_I}}
\chi(\omega_P;\omega_S,\omega_I)\sqrt{\Delta \omega}.
\end{eqnarray}
We are interested in the regime far below threshold where the biphoton production rate $\kappa$ is much smaller than the cavity damping rate,
\begin{equation}
\label{kapp} \kappa \ll \gamma,
\end{equation}
and the mean photon number in the resonator therefore close to zero. Since the mean time interval between biphoton emission events is large compared
to the cavity damping time, the resonator can be assumed to be empty before each emission event. For this case, we can perform a perturbative
treatment of the nonlinear interaction, assuming that at the initial time $t=0$, the combined signal-idler field is in the vacuum state, described by
$|0\rangle = |0\rangle_S\otimes|0\rangle_I $. In the following we rely on the ideas developed for the theory of spontaneous parametric
down-conversion \cite{hong,ghosh,mandel}. By expanding the formal solution of the time-dependent Schr\"odinger equation, up to a normalization
factor, the state vector describing the combined signal-idler field at a time $t=\Delta t\ll \kappa^{-1}$ is found to be $|\psi(\Delta t) \rangle
\propto |0\rangle + |\tilde{\psi}(\Delta t) \rangle$ with
\begin{equation}
\label{psi-tilde} |\tilde{\psi}(\Delta t) \rangle = \frac{1}{i\hbar} \int_0^{\Delta t} dt\,H_{int}(t) |0\rangle.
\end{equation}
After substituting Eq. (\ref{H-int1}) into Eq. (\ref{psi-tilde}), integration with respect to $t$ yields
\begin{eqnarray} \label{psi-tilde1}
\lefteqn{ |\tilde{\psi}(\Delta t) \rangle =\sum_{m=-\infty}^{\infty} \int_{-\infty}^{\infty}d\Omega
\frac{\alpha\sqrt{\gamma}}{\frac{\gamma}{2}-i\Omega}
\int_{-\infty}^{\infty}d\Omega^{\prime}F_m (\Omega,\Omega^{\prime})}\nonumber\\
&& \times\,\Delta t \;{\rm sinc} \left[\frac{1}{2}(m \Delta \omega + \Omega + \Omega^{\prime})\Delta t\right]
  {\rm e}^{ \frac{i}{2}(m\Delta\omega + \Omega + \Omega^{\prime})\Delta t}\nonumber\\
&& \times\,a^{\dag} (\omega_m+\Omega) b^{\dag} (\omega_I+\Omega^{\prime})|0\rangle,
\end{eqnarray}
where we used the sinc-function defined as ${\rm sinc}(z) = \frac{\sin z}{z}$ with ${\rm sinc}(0)=1$.

The nonnormalized vector $|\tilde{\psi}(\Delta t) \rangle $ refers to the state of the radiation field on the condition that a signal-idler photon
pair has been produced during the time interval $\Delta t$. Since this condition applies with the probability $ \langle \tilde{\psi}(\Delta t)
|\tilde{\psi}(\Delta t)\rangle$, the biphoton production rate is given by
\begin{equation}
\label{kappa} \kappa = \frac{1}{\Delta t}\langle \tilde{\psi}(\Delta
t) |\tilde{\psi}(\Delta t)\rangle.
\end{equation}
Applying the commutation relations Eqs.~(\ref{comm1}) and
(\ref{comm3}), we get from Eq.~(\ref{psi-tilde1})
\begin{eqnarray} \label{kappa1}
\lefteqn{ \langle \tilde{\psi}(\Delta t) |\tilde{\psi}(\Delta t)\rangle =
(\Delta t)^2 \sum_{m=-\infty}^{\infty} \int_{-\infty}^{\infty}d\Omega\,
\frac{|\alpha|^2\gamma}{(\frac{\gamma}{2})^2+\Omega^2}}\\
&&\times\int_{-\infty}^{\infty}d\Omega^{\prime}\,
|F_m (\Omega,\Omega^{\prime})|^2\; {\rm sinc}^2\!\!\left[\frac{1}{2}(m \Delta\omega + \Omega +
\Omega^{\prime})\Delta t\right].\nonumber
\end{eqnarray}
Because of the properties of the sinc-function, the integral is dominated by the region $m \Delta\omega + \Omega + \Omega^{\prime}\leq \pi/\Delta t$.
If $\Delta t$ is sufficiently large and  $F_m(\Omega,\Omega^{\prime})$ is a slowly varying function of $\Omega^{\prime}$ in this region, the latter
function can be replaced by its value at $\Omega^{\prime}= - m\Delta \omega - \Omega$ \cite{mandel}. Using the relation $\int_{-\infty}^{\infty}{\rm
sinc}^2(z) dz=\pi$, the integration with respect to $\Omega^{\prime}$ is then readily performed, and we obtain
\begin{eqnarray}
\label{kappa2}
\lefteqn{{\langle \tilde{\psi}(\Delta t) |\tilde{\psi}(\Delta
t)\rangle}=} \\
&&   2 \pi \,|\alpha|^2 \Delta t
 \sum_{m=-\infty}^{\infty}
\int_{-\infty}^{\infty}d\Omega\, \frac{\gamma}{(\frac{\gamma}{2})^2+\Omega^2}\, |\Phi_m (\Omega)|^2,\nonumber
\end{eqnarray}
where
\begin{equation}
\label{phim} \Phi_m(\Omega)= F_m(\Omega, - m\Delta \omega - \Omega).
\end{equation}
For further evaluation, we need to specify the function $\Phi_m(\Omega)$. Using  Eqs.~(\ref{k}), (\ref{kI}), and (\ref{kS}) with $\omega_m=\omega_S+
m\Delta \omega$, Taylor expansion around the central frequencies $\omega_S$ and $\omega_I$ yields
\begin{eqnarray}
\label{Dk} k_P- k_{S,m}(\Omega)-k_I(-m\Delta \omega\!-\!\Omega)\approx(m\Delta\omega\!
 +\! \Omega) \frac{\tau_0}{l},\qquad
\end{eqnarray}
where we introduced the time constant
\begin{eqnarray} \label{tau0}
\tau_0= \frac{l}{c}\left(n_I + \left.\omega_I\frac{\partial n_i}
{\partial \omega}\right|_{\omega=\omega_I}\!\!\!\! -n_S -
\left.\omega_S\frac{\partial n_s}{\partial \omega}\right|_{\omega=\omega_S}\right).\qquad
\end{eqnarray}
The latter is equivalent to
\begin{equation}
\label{tau01} \tau_0= \frac{l}{v_{g,I}}- \frac{l}{v_{g,S}}
\end{equation}
and describes the difference between the transit times of a signal and idler photon through a crystal of length $l$, originating  from the difference
in the signal and idler group velocities, ${v_{g,S}}$ and ${v_{g,I}}$, defined by Eq.~(\ref{vg}) and by the corresponding equation for the idler
wave, respectively. Since in any real experiment $|{v_{g,S}}- {v_{g,I}}| \ll {v_{g,S}}$, it follows that $|\tau_0|\ll(\Delta\omega)^{-1}$ where  we
used Eq.~(\ref{T}). Hence, we are considering a parameter range in this problem that is characterized by the combined inequality
\begin{eqnarray}
\label{inequality1} \kappa \ll \gamma \ll\ \Delta\omega \ll |\tau_0|^{-1},
\end{eqnarray}
where Eqs.~(\ref{goodcavity}) and (\ref{kapp}) have been incorporated \cite{parameters}. Assuming a constant nonlinear susceptibility within the
bandwidth given by $|\tau_0|^{-1}$, we find from Eqs.~(\ref{phim}), (\ref{Dk}), and (\ref{F-m}) that
\begin{eqnarray}
\label{phim1} \Phi_m(\Omega) \approx \frac{1}{l}  \int_{-l}^0 dx\; {\rm e}\,^{i
(m\Delta\omega+\Omega)\frac{\tau_0}{l}x}.
\end{eqnarray}

In order to determine the rate of biphoton generation, we have to insert Eq.~(\ref{phim1}) into Eq.~(\ref{kappa2}) where the integration with respect
to $\Omega$ is effectively restricted to the cavity bandwidth $\gamma$ with $\gamma\ll\Delta\omega$. Hence, we neglect $\Omega$ compared to
$m\Delta\omega$ in Eq.~(\ref{phim1}) for $m\neq 0$. Moreover, Eq.~(\ref{inequality1}) implies $|\Omega\tau_0|\ll 1$ in the relevant interval
$|\Omega| \lesssim \gamma$ and therefore $\Phi_0(\Omega)\approx 1$. After integration with respect to $x$, we get the approximation
\begin{equation}
\label{phim2} \Phi_m(\Omega)\approx {\Phi}_m (0)= {\rm sinc} \left(m
\Delta\omega \frac{\tau_0}{2}\right) {\rm e}^{-i m \Delta\omega
\frac{\tau_0}{2}}
\end{equation}
which can be used in connection with Eq.~(\ref{kappa2}), (\ref{kappa}), and (\ref{alpha}) to determine the rate of biphoton generation
\begin{equation}
\label{kappa3} \kappa=  \left(\frac{\chi |E_p| }{4 \epsilon_0c
A}\right)^2\frac{\omega_S \omega_I}{n_S n_I} \;\Delta\omega
\sum_{m={-\infty}}^{\infty} {\rm sinc}^2 \left(m \Delta\omega
\frac{\tau_0}{2}\right).
\end{equation}
Further simplification is possible if we transform the sum into an integral with respect to $z=m\Delta\omega\tau_0/2$, introducing the positive
increment $dz= \Delta\omega|\tau_0|/2$ where $dz\ll 1$ because of Eq.~(\ref{inequality1}). We then arrive at the expression
\begin{equation}
\label{kappa4} \kappa \approx \left(\frac{\chi |E_p| }{4 \epsilon_0c
A}\right)^2\frac{\omega_S \omega_I}{n_S n_I} \frac{2\pi}{|\tau_0|}
\end{equation}
that does not depend on the properties of the resonator because the mean photon number in our cavity is approximately zero and, in addition, an
associated idler mode exists for each signal mode which meets the requirement for energy conservation.

The presented perturbative treatment relies on the condition $|\tau_0|\ll \Delta t\ll\kappa^{-1}$ so that the wave function $|\tilde{\psi}(\Delta
t)\rangle$ in Eq.~(\ref{psi-tilde1}) and the approximation leading to Eq.~(\ref{kappa2}) are valid simultaneously. The normalized vector
$|\psi\rangle=(\kappa \Delta t)^{-1/2}|\tilde{\psi}(\Delta t)\rangle$ can be denoted as the biphoton wave function since it represents the state of
the radiation field on the condition that exactly one signal-idler photon pair is present.

A more convenient representation of the biphoton wave function is obtained if the sinc-function in Eq.~(\ref{psi-tilde1}) is replaced by
$2\pi\delta(m\Delta\omega + \Omega+\Omega^{\prime})$ according to the standard procedure \cite{ghosh}. Mathematically, this corresponds to the limit
$\Delta t \rightarrow \infty$, implying $\kappa \rightarrow 0$. Then, the integration with respect to $\Omega^{\prime}$ can be performed immediately.
In analogy to the expression given in Ref.~\cite{ghosh}, we obtain the biphoton wave function
\begin{eqnarray} \label{psi}
\lefteqn{|\psi\rangle  =  {\cal N} \sum_{m=-\infty}^{\infty}
\int_{-\infty}^{\infty}d\Omega\;
\frac{\Phi_m(\Omega)}{\frac{\gamma}{2}-i\Omega}}\nonumber\\
&& \times\,a^{\dag} (\omega_S + m\Delta\omega + \Omega)\, b^{\dag}
(\omega_I-m\Delta\Omega-\Omega)|0\rangle, \qquad
\end{eqnarray}
where  $\Phi_m(\Omega)$ is given by Eq.~(\ref{phim1}) and ${\cal N}$ is a normalization constant. The explicit value of ${\cal N}$~\cite{footnote} is
not important as long as only normalized quantities characterizing the radiation field are considered. Eq.~(\ref{psi}) clearly reveals the
frequency-entanglement between the signal and idler photon and will serve as our basic equation to determine the properties of the emitted radiation.

\section{Properties of the emitted radiation}
\subsection{Output spectra of the signal and idler field}
First we investigate the output spectra $S_{S}(\omega)$ and $S_{I}(\omega)$ of the signal and  idler field, defined as
\begin{eqnarray}
\label{spectrum} S_{S/I}(\omega) = \frac{1}{2\pi}\int_{-\infty}^{\infty}
d\tau \;G_{S/I}^{(1)}(\tau)\;{\rm e}^{i\omega\tau},
\end{eqnarray}
where $G_{S}^{(1)}(\tau)$ and $G_{I}^{(1)}(\tau)$ are the first-order temporal correlation functions of the respective fields outside the resonator.
In the following, it is convenient to use the Heisenberg picture and to start from the expression
\begin{eqnarray}
\label{corr1} G_{S/I}^{(1)}(\tau) &=&\langle \psi
|\,E_{S/I}^{(-)}(x,t)\,E_{S/I}^{(+)}(x,t+\tau)\,|\psi\rangle.
\end{eqnarray}
Here, $|\psi\rangle$ is the time-independent biphoton wave function given by Eq.~(\ref{psi}) and $E_S^{(+)}= E_S^{(-)\dag}$ and
$E_I^{(+)}=E_I^{(-)^\dag}$ denote the positive-frequency parts of the time-dependent operators of the electric fields in free space, given by
Eqs.~(\ref{sig-out}) and (\ref{id1}). Let us first determine the spectrum of the idler field. Considering
\begin{figure}[t!]
\center{\includegraphics[scale=0.8,draft=false]{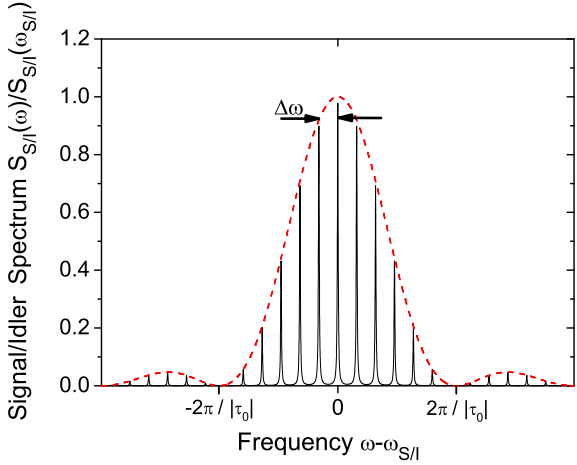}} \caption{(Color online) Schematic plot of the normalized signal and idler output
spectra. The width of the Lorentzian peaks is determined by the cavity damping rate $\gamma$, and they are separated by the free spectral range
$\Delta\omega=2\pi/T$. The envelope, described by the sinc-function in Eq. (\ref{SI}), yields the total spectral width $2\pi/|\tau_0|$.}
\end{figure}
\begin{equation}
\label{comm} b(\omega_I +\Omega^{\prime})b^{\dag}
(\omega_I-m\Delta\omega-\Omega)|0\rangle=\delta(\Omega^{\prime}+m\Delta\omega+\Omega)|0\rangle,
\end{equation}
due to Eq.~(\ref{comm1}), we obtain from Eqs.~(\ref{id1}) and (\ref{psi})
\begin{eqnarray}
\label{help} \lefteqn {E_I^{(+)}(x,t)|\psi\rangle \propto \sum_{m=-\infty}^{\infty}
\int_{-\infty}^{\infty}d\Omega\,\frac{\Phi_m(\Omega)}{\frac{\gamma}{2}-i\Omega}} \\
&&\times\, {\rm e}^{i(\omega_I-m\Delta\omega-\Omega)\left(\frac{x}{c}-t\right)}
a^{\dag} (\omega_S + m\Delta \omega + \Omega) |0\rangle.\nonumber
\end{eqnarray}
By taking the inner product of  $E_I^{(+)}(x,t+\tau)|\psi\rangle$
and $E_I^{(+)}(x,t)|\psi\rangle$ and applying the commutation
relation Eq.~(\ref{comm3}), we find
\begin{eqnarray}
\label{GI} G_I^{(1)}(\tau) \propto\sum_{m=-\infty}^{\infty}
\int_{-\infty}^{\infty}d\Omega\,
\frac{|\Phi_m(\Omega)|^2}{\left(\frac{\gamma}{2}\right)^2+\Omega^2}
{\rm e}^{-i(\omega_I-m\Delta\omega-\Omega)\tau}.\nonumber\\
\end{eqnarray}
In analogy to the derivation of Eq.~(\ref{kappa3}), the dependence
of $\Phi_m$ on $\Omega$ can be neglected within the relevant
bandwidth determined by $\gamma$. After inserting Eq.~(\ref{phim2})
into Eq.~(\ref{GI}), the Fourier transform yielding the idler
spectrum according to Eq.~(\ref{spectrum}) is readily performed. The
signal spectrum can be determined in a completely analogous way, and
we finally arrive at the relation
\begin{equation}
\label{SI} S_{S/I}(\omega)\propto \sum_{m={-\infty}}^{\infty} \frac{
{\rm sinc}^2 \left(m \Delta\omega \frac{\tau_0}{2}\right)}
{\left(\frac{\gamma}{2}\right)^2+(\omega_{S/I}-m\Delta\omega -
\omega)^2}.
\end{equation}
Eq. (\ref{SI}) indicates that both the signal and idler spectrum are
composed of Lorentzians of halfwidth $\gamma$ centered at
frequencies $\omega_{S/I}-m\Delta\omega$ with $m=0,\pm1,\ldots$
where the respective spectral envelopes are determined by the
sinc-function in the nominator. From Fig.~2 it becomes obviouos that
the frequency bandwidth of the signal and idler photons is
characterized by $|\tau_0|^{-1}$. Their temporal uncertainty is thus
equal to the modulus of the time constant $\tau_0$ introduced in
Eq.~(\ref{tau0}) and resulting from the phase-matching conditions.
Even though the idler wave is not resonant, the longitudinal mode
structure of the resonator is revealed in the idler spectrum due to
the frequency entanglement between the signal and idler photon which
arises from the interaction underlying the biphoton generation
process.

\subsection{Signal-idler cross-correlations}
The coincidence rate for detecting an idler photon at time $t$ and a signal photon
at time $t+\tau$, both at equal distance from the end facet of the
crystal, is proportional to the temporal correlation function
\begin{eqnarray}
\label{G2} \lefteqn{G_{IS}^{(2)}(\tau)=}\\
 && \langle \psi
|E_{I}^{(-)}(x,t)E_{S}^{(-)}(x,t+\tau)E_{S}^{(+)}(x,t+\tau)E_{I}^{(+)}(x,t)|\psi\rangle,\nonumber
\end{eqnarray}
where again $E_S^{(+)}$ and $E_I^{(+)}$ are the free-field operators
defined by Eqs.~(\ref{sig-out}) and (\ref{id1}). With the explicit
expression for the biphoton wave function $|\psi\rangle$, given by
Eq.~(\ref{psi}), we get
\begin{eqnarray}
\label{help1}E_{S}^{(+)}(x,t+\tau)E_{I}^{(+)}(x,t)|\psi\rangle
\;\propto \;\;
{\rm e}\,^{i\left[(\omega_S+\omega_I)\left(\frac{x}{c}-t\right)-\tau\,\omega_S\right]\quad\nonumber}\\
\times\sum_{m=-\infty}^{\infty} {\rm e}^{-im \Delta\omega\tau}
\int_{-\infty}^{\infty}d\Omega\,
\frac{\Phi_m(\Omega)}{\frac{\gamma}{2}-i\Omega} {\rm e}^{-i
\Omega\tau}|0\rangle,\qquad
\end{eqnarray}
where Eq.~(\ref{comm}) and the corresponding relation for the signal modes
\begin{equation}
\label{comm4} a(\omega_S +\Omega^{\prime})a^{\dag}(\omega_S+m\Delta\omega+\Omega)|0\rangle=
\delta(\Omega^{\prime}-m\Delta\omega-\Omega)|0\rangle,
\end{equation}
following from Eq.~(\ref{com-a}), have been applied. According to
Eq.~(\ref{G2}), the correlation function $G_{IS}^{(2)}(\tau)$ is
proportional to the squared norm of the Hilbert vector on the
right-hand side of Eq.~(\ref{help1}). It can be determined from the
explicit expression for $\Phi_m(\Omega)$, given by
Eq.~(\ref{phim1}), together with the integral identities
\begin{eqnarray}
\label{int} - \frac{1}{\pi } \int_{-\infty}^{\infty}d\Omega\,
\frac{{\rm e}^{-i\Omega t}} {\frac{\gamma}{2}-i\Omega }= \left \{
\begin{array}{ll}
 0 \;\; & \mbox{if $ t < 0$} \\
1\;\; & \mbox{if $ t = 0$}\\
 2{\rm e}^{-\frac{\gamma}{2}t}\;\; & \mbox{if $ t > 0.$}
 \end{array}
\right.
\end{eqnarray}
First, it is important to observe that
\begin{eqnarray}
\label{G4} G_{IS}^{(2)}(\tau)=0\qquad{\rm if}\; &&\mbox{ $ \tau +
\frac{\tau_0}{2} < -\frac{|\tau_0|}{2}$}
\end{eqnarray}
\begin{figure}[t!]
\center{\includegraphics[scale=0.7,draft=false]{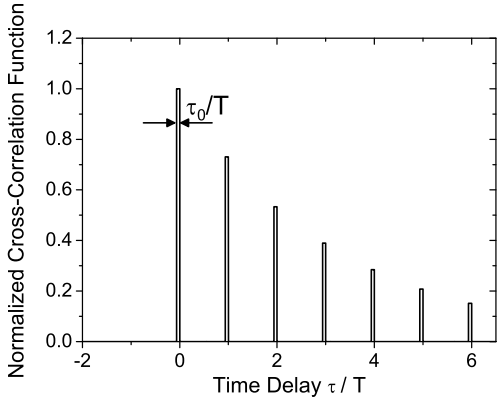}}
\caption{Schematic representation of the normalized second-order
signal-idler cross-correlation function $G_{IS}^{(2)}(\tau)$
according to Eq. (\ref{G6}) for $\tau_0>0$ and
$\gamma/\Delta\omega=0.05$. If the time delay $\tau$ is equal to
zero or multiples of the cavity round-trip time
$T=2\pi/\Delta\omega$, the function exhibits pronounced peaks which
decay with the cavity damping time $\gamma^{-1}$.}
\end{figure}which is equivalent to $\tau < -\tau_0$ for $\tau_0>0$ and  $\tau
<0$ for $\tau_0<0$, respectively. Mathematically, Eq.~(\ref{G4}) is
due to  the expression $\Omega t=\Omega(\tau-\frac{x}{l}\tau_0)$
which results from inserting Eq.~(\ref{phim1}) into (\ref{help1})
and is negative in the given case for any $x$ inside the crystal,
i.e. for $-l\leq x \leq 0$. Therefore, the upper line of
Eq.~(\ref{int}) applies. Physically, the correlation function
vanishes because the arrival time of the signal photon in a photon
pair can precede the arrival time of the corresponding idler photon
at most by a time interval that is within the temporal uncertainty
interval $|\tau_0|$ inherent in the biphoton generation process.

On the other hand, the presence of the resonator allows a signal
photon to be detected considerably later than the associated idler
photon since the signal photon may bounce back and forth between the
resonator mirrors repeatedly before leaving the resonator. For time
delays $\tau$ outside the limits of Eq.~(\ref{G4}), we approximate
$\Phi_m(\Omega)$ by Eq.~(\ref{phim2}) where the integration with
respect to $x$ has been performed and the dependence on $\Omega$ has
been neglected. After inserting Eq.~(\ref{phim2}) into
Eq.~(\ref{help1}), the damping of the cavity gives rise to a factor
proportional to ${\rm e}^{-\frac{\gamma}{2}\tau}$ for
$G_{IS}(\tau)\neq 0$ which follows from the third line of
Eq.~(\ref{int})~\cite{footnote1}. Since $G_{IS}^{(2)}(\tau)= \|
E_{S}^{(+)}(x,t+\tau)E_{I}^{(+)}(x,t)|\psi\rangle\|^2$, we finally
get from Eq.~(\ref{help1}) the approximate result
\begin{eqnarray}
\label{G5}
G_{IS}^{(2)}(\tau)&\propto& {\rm e}^{-\gamma \tau} \left|
\sum_{m=-\infty}^{\infty}{\rm sinc} \left(m \Delta\omega \frac{\tau_0}{2}\right){\rm
e}^{-im\Delta\omega\left(\tau+\frac{\tau_0}{2}\right)}
 \right|^2\nonumber\\
&&\mbox{if $\;\; \tau + \frac{\tau_0}{2} \geq
-\frac{|\tau_0|}{2}$}.
\end{eqnarray}
According to Eq.~(\ref{tau01}), the sign of $\tau_0$ is determined
by the relation between the signal and idler group velocities, i.e.
$\tau_0 > 0$ for $v_{g,S}>v_{g,I}$ and $\tau_0 < 0$ for
$v_{g,S}<v_{g,I}$. A numerical evaluation reveals that
Eq.~(\ref{G5}) describes a decaying periodic function with peaks of
width $|\tau_0|$ centered at $\tau= j\, T-\tau_0/2$ where
$j=0,1,\ldots$ and $T=2\pi/\Delta\omega $ is the cavity round-trip
time. The time shift $ -\tau_0/2$ of the peaks arises since for
$\tau_0>0$ the time needed by the center of the signal wave packet
to travel from the middle of the crystal to its end facet is by the
amount $\tau_0/2$ shorter than the time needed by the center of the
idler wave packet to cover the same distance while the opposite
holds for $\tau_0 < 0$.

Transforming the sum in Eq.~(\ref{G5}) into an integral, the
expression for the signal-idler cross-correlation function  can be
further approximated, in analogy to the procedure applied to derive
Eq.~(\ref{kappa4}). Introducing $z=m\Delta\omega\tau_0/2$ and $dz=
\Delta\omega\tau_0/2$, we find that $G_{IS}^{(2)}(\tau) \propto {\rm
e}^{-\gamma \tau}I^2$ with $I=\int_{-\infty}^{\infty}{\rm sinc}(z)
\cos(az)\,dz$ since ${\rm sinc}(z)$ is an even function of $z$.
Because of the periodicity of the cos-function, the parameter $a$
can be written as
$a=1+2\left[\tau-\left(\left[\frac{\tau}{T}\right]+1\right)T\right]/\tau_0$
where $\left[\frac{\tau}{T}\right]$ denotes the largest integer that
does not exceed $\tau/T$. Since  $I=\pi$ for $|a|<1$ and $I=0$ for
$|a|>1$, Eq.~(\ref{G5}) takes a simple form that can be combined
with Eq.~(\ref{G4}) to yield the compact approximate representation
\begin{eqnarray}
\label{G6} G_{IS}^{(2)}(\tau)\propto \sum_{j=0}^\infty\left\{\begin{array}{ll}
  {\rm e}^{-\gamma j T}\quad\mbox{if $\;\;
  \left|\tau-j\,T+ \frac{\tau_0}{2}\right|\leq
  \frac{|\tau_0|}{2}\quad$}\\
   0\qquad\quad\mbox{else,}
\end{array}
\right.
 \end{eqnarray}
where $ j= \left|\left[\frac{\tau}{T}\right]\right|$ (see Fig.~3).
Here, we took into account that due to Eq.~(\ref{inequality1}) cavity damping is
negligible during the time interval $|\tau_0|$, i.e. ${\rm exp}(-\gamma\tau_0)\approx 1$.

\section{Discussion and conclusions}
\begin{figure}[b!]
\center{\includegraphics[scale=0.7,draft=false]{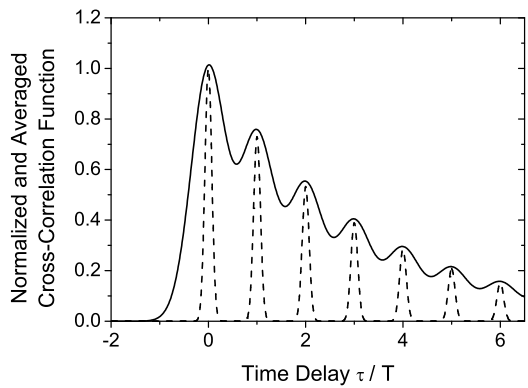}}
\caption{Time-averaged second-order signal-idler cross-corre\-lation
function $\overline{G_{IS}}^{(2)}(\tau)$ for
$\gamma/\Delta\omega=0.05$. The ratio between the resolution time
$\Delta T$ and the cavity round-trip time $T=2\pi/\Delta\omega$ is
assumed as $\Delta T/T= 0.02$ (dashed line) and $\Delta T/T= 1$
(solid line), respectively.}
\end{figure}
In a real experiment, the sharp peaks of the function
$G_{IS}^{(2)}(\tau)$ will be broadened due to the finite resolution
time of the detector setup. When the latter is taken into account by
performing the convolution with respect to a Gaussian function,
Eq.~(\ref{G6}) yields the time-averaged cross-correlation function
\begin{equation}
\label{G7} \overline{G_{IS}}^{(2)}(\tau)\propto
  \sum_{j=0}^{\infty}{\rm
  exp}\left[-\gamma j T-\frac{4(jT-\tau)^2}{(\Delta T)^2}\right],
\end{equation}
where $\Delta T$ characterizes the effective resolution time and
where we have assumed $\Delta T \gg |\tau_0|$. The resulting
averaged function is plotted in Fig. 4 for two different values of
$\Delta T$. A second-order cross-correlation function showing the
behavior of the solid line in Fig.~4 has recently been measured in
our group~\cite{scholz}, and the results have been found to be in
excellent agreement with the predictions derived from
Eqs.~(\ref{G4}) and (\ref{G5}).

We still note that for spontaneous parametric down-conversion in a
double-resonant cavity the signal-idler cross-correlation function
has also been found to exhibit a comb-like structure which is,
however, symmetric with respect to the time
delay~\cite{lu2,wang,kukl2}. The effect has been explained by
applying the concept of mode-locking to the frequency-entangled
biphoton state, pointing out that due to the large coherence time of
the pump, photon pairs with different frequencies have a common
phase and form a coherent superposition~\cite{lu2}.

To summarize, we performed a theoretical investigation of biphoton
generation by spontaneous parametric down-conversion in a
single-resonant OPO far below threshold. We derived analytical
expressions for the rate of biphoton generation, for the output
spectra of the signal and idler fields, as well as for the second
order signal-idler cross correlation function. Our investigations
provide the theoretical background for explaining the results of a
recent experiment \cite{scholz}, where stable continuous operation
of a single-resonant OPO far below threshold has been demonstrated.

\section*{Acknowledgments}
This work was supported by Deutsche Forschungsgemeinschaft DFG, grant BE~2224/5.
M. Scholz acknowledges funding by Deutsche Telekom Stiftung.


\begin{thebibliography}{99}
\bibitem{footnote}  For completeness, we state the formal result
$ {\cal N}^2 = |\alpha|^2 2\pi\gamma/[\kappa \delta(0)]$ which follows with the help of
Eqs. (\ref{comm3}), (\ref{kappa}) and (\ref{kappa2}). The delta function arises since
   $\langle 0| b (\omega)b^{\dag} (\omega)|0\rangle = \delta(0)$ because of Eq.~(\ref{comm1}).
   Note that $\kappa\rightarrow 0$.
\bibitem{parameters} In a typical experiment~\cite{scholz}, the relevant parameters are
given by $\kappa \sim 0.1$~MHz, $ \gamma \sim 10$~MHz, $\Delta\omega
\sim 1$~GHz, and $|\tau_0|^{-1} \sim 150$~GHz.
\bibitem{footnote1} Since a mathematically exact treatment of the discontinuity is physically irrelevant,
we replace the value of the integral in the second line of Eq. (\ref{int}) by the value of the third line.
\end{thebibliography}
\end{document}